\documentclass[conference]{IEEEtran}
\IEEEoverridecommandlockouts

\usepackage{cite}
\usepackage{amsmath,amssymb,amsfonts}
\usepackage{algorithmic}
\usepackage{graphicx}
\usepackage{textcomp}
\usepackage{xcolor}
\usepackage[hyphens]{url}
\usepackage{fancyhdr}
\usepackage{hyperref}

\usepackage{booktabs}
\usepackage{multirow}
\usepackage{listings}
\usepackage{tabularx}
\usepackage{subcaption}
\usepackage{booktabs}
\usepackage{array} 
\usepackage{xspace}

\newcommand{\aname}{TOM\xspace}
\newcommand{\figref}[1]{Figure~{\ref{#1}}}

\def\BibTeX{{\rm B\kern-.05em{\sc i\kern-.025em b}\kern-.08em
    T\kern-.1667em\lower.7ex\hbox{E}\kern-.125emX}}
\begin{document}

\pdfpagewidth=8.5in
\pdfpageheight=11in




\newcommand{\iscasubmissionnumber}{469}
\title{TOM: A Ternary Read-only Memory Accelerator for LLM-powered Edge Intelligence}
\pagenumbering{arabic}

\author{
    Hongyi Guan\textsuperscript{1}, 
    Yijia Zhang\textsuperscript{2,3}, 
    Wenqiang Wang\textsuperscript{2,3}, 
    Yizhao Gao\textsuperscript{4}, 
    Shijie Cao\textsuperscript{1,$\ast$}, 
    Chen Zhang\textsuperscript{2}, 
    Ningyi Xu\textsuperscript{2,3}
    \\
    \textsuperscript{1}Microsoft Research, \textsuperscript{2}Shanghai Jiao Tong University, \\
    \textsuperscript{3}Huixi Intelligence, \textsuperscript{4}The University of Hong Kong \\
    
    \small{hongyi.guan.thu@outlook.com, \{zhangyijia, wangwq20, chenzhang.sjtu, xuningyi\}@sjtu.edu.cn,} \\
    \small{yizhao@connect.hku.hk, caoshijie0501@gmail.com}
    \thanks{$\ast$ Corresponding author.}
}

\maketitle
\thispagestyle{plain}
\pagestyle{plain}



\begin{abstract}

The deployment of Large Language Models (LLMs) for real-time intelligence on edge devices is rapidly growing.
However, conventional hardware architectures face a fundamental memory wall challenge, where limited on-device memory capacity and bandwidth severely constrain the size of deployable models and their inference speed, while also limiting on-device adaptation.
To address this challenge, we propose TOM, a hybrid ROM-SRAM accelerator co-designed with ternary quantization, which balances extreme density with on-device tunability.
TOM exploits the synergy between ternary quantization and ROM to achieve extreme memory density and bandwidth, while preserving flexibility through a hybrid ROM-SRAM architecture designed for QLoRA-based tunability.
Specifically, we introduce: (1) a sparsity-aware ROM architecture that synthesizes ternary weights as standard-cell logic, eliminating area overhead from zero-valued bits; (2) a distributed processing architecture that co-locates high-density ROM banks with flexible SRAM-based QLoRA adapters and compute units; and (3) a workload-aware dynamic power gating scheme that exploits the logic-based nature of ROM to power down inactive banks, minimizing dynamic energy consumption.
TOM achieves an inference throughput of 3,306 TPS using BitNet-2B model, demonstrating its effectiveness in delivering real-time, energy-efficient edge intelligence.

\end{abstract}

\section{Introduction}

Large Language Models (LLMs) are increasingly being deployed on edge devices such as smartphones, autonomous vehicles, robots, and AR glasses~\cite{qu2025mobile}. This trend is driven by the demand for real-time, context-aware intelligence. Deploying LLMs locally offers compelling benefits, including reduced end-to-end latency, enhanced privacy, reliable offline operation, and reduced dependency on cloud infrastructure.

However, deploying LLMs on edge devices presents stringent challenges in memory, compute, and power efficiency. Among these, memory stands out as the dominant bottleneck, as limited on-device storage and bandwidth severely constrain both model size and inference performance.
As the scale and diversity of edge AI applications continue to expand, the demand for more powerful and efficient LLMs on edge devices is rapidly increasing. Future edge intelligence workloads are expected to require 1000 TPS, driven by scenarios involving multimodal input and real-time reasoning~\cite{li2024large}.
Addressing this challenge calls for \textit{model-hardware co-design} to enable efficient edge intelligence.

On the model side, low-bit quantization has emerged as the most practical strategy to reduce the footprint of LLMs. While 4-bit weight quantization has become pervasive, both academia and industry are actively exploring advancements toward 2-bit and even 1-bit quantization. In particular, ternary quantization, with weights constrained to \(\{ -1, 0, +1 \}\), strikes a favorable balance between compression and accuracy. Notably, ternary quantization is a game-changer for hardware design because it eliminates multiplications and leaves only additions. This simplicity dramatically reduces computational complexity and energy consumption.
These advantages make ternary LLMs highly attractive for efficient edge deployment.

On the hardware front, various architectural innovations have emerged to address memory bottlenecks. These include Processing-in-Memory (PIM)~\cite{kim2022overview}, distributed SRAM systems, e.g., Cerebras, Dojo, Groq~\cite{lie2022cerebras,talpes2022dojo,abts2022groq}, and 3D-stacked DRAM~\cite{yang2024enabling}. While these approaches offer benefits like reduced data movement or high bandwidth, they often come with significant trade-offs in scalability, power, or cost, making them challenging for edge deployment. Critically, these prior solutions struggle to simultaneously achieve the high density and high bandwidth required for efficient LLM inference on highly constrained edge platforms. This gap highlights the need for a novel hardware architecture tailored for quantized LLMs at the edge.

Beyond conventional solutions, we propose a strategic two-tier architecture for on-device intelligence. We leverage Read-Only Memory (ROM) as an energy-efficient 'knowledge foundation', storing the quantized base model which is updated on a hardware lifecycle (e.g., yearly). This is motivated by the trend of smaller base models saturating, where future value will stem from adaptation \cite{gunter2024apple, tina}. Therefore, an SRAM-based 'flexibility layer' provides crucial runtime adaptability, housing QLoRA modules \cite{hu2022lora,qlora,lora-qaf} for frequent model tuning and, critically, for correcting potential base model errors or post-fabrication defects. This hybrid approach preserves the ROM's core efficiency while addressing reliability and flexibility concerns.

Despite its promising potential, conventional ROM technologies remain insufficient to meet the stringent storage and bandwidth requirements of edge intelligence. 
To fully exploit ROM for LLMs, we argue that model–hardware co-design is indispensable. In particular, low-bit quantization, especially ternary, opens new opportunities to drastically increase effective memory density while simplifying hardware~\cite{zhou2024survey}. However, co-designing ROM with ternary LLMs introduces several unique challenges. First, ternary weights inherently exhibit high sparsity, raising questions about how to encode and exploit this sparsity to further boost ROM density without incurring complex access patterns. Second, edge LLMs often adopt heterogeneous quantization schemes, where linear and attention computation may use different precisions. Designing ROM-based accelerators that efficiently support such mixed-precision computation, while fully leveraging ROM’s ultra-high bandwidth, is non-trivial. Finally, power consumption remains a critical constraint for battery-powered devices. Although ROM’s low-leakage nature is advantageous, integrating it into a high-performance computing fabric demands careful architectural design to minimize dynamic energy during frequent weight accesses.

To address the above challenges, we propose \textbf{TOM}, a \textbf{T}ernary Read-\textbf{O}nly \textbf{M}emory accelerator to boost the efficiency LLM inference on edge scenarios. First, we exploit the bit-level sparsity of ternary weights in LLM and design a sparsity-aware ROM implementation. Typically, we synthesize ROM as a combinational logic circuit to output one-value bits given input address. Compared to standard ROM implementation that use rigid grid-like physical structure using either via-programmed or mask-programmed techniques, our sparsity-aware ROM can achieve much higher memory density. 
Second, \aname employs a distributed processing architecture that co-locates computation with memory, organizing parallel Processing Lanes, each with local ROM and compute units to maximize bandwidth and minimize data movement. A centralized global controller and reduction tree coordinate the lanes and aggregate partial results, enabling efficient and scalable transformer inference on-chip. For each compute unit, it contains a GEMV unit with heterogeneous quantization support: Ternary $\times$ FP8 or FP8 $\times$  FP8, which share identical adder tree to save area. Finally, since our ROM design is based on standard cell logic, \aname has the ability to perform aggressive power gating to power down inactivated layers' weights during inference, which greatly improves energy efficiency in edge devices. 

We evaluate \aname using the BitNet-2B model to demonstrate the effectiveness of our model-hardware co-design approach. Our evaluation shows that TOM achieves a peak inference throughput of 3,306 TPS, outperforming a high-end NVIDIA A100 GPU by up to 465x. This remarkable performance stems from our novel sparsity-aware ROM, which achieves a state-of-the-art storage density of 15.0 MB/mm² , and our workload-aware dynamic power gating, which slashes total chip power by nearly 80\% to a mere 5.33W. These results validate TOM as a highly efficient architecture capable of enabling the next generation of real-time intelligence on edge devices.

\noindent Our main contributions are as follows:

\begin{itemize}
    \item We explore the co-design of ternary quantization and read-only memory (ROM) for efficient edge LLM deployment, showing that their synergy enables compact, high-performance, and energy-efficient inference accelerators.

    \item We propose \aname, a novel accelerator architecture featuring three key techniques to achieve extreme memory density and bandwidth: (1) a sparsity-aware ROM that synthesizes ternary weights into standard-cell logic to maximize density , (2) a distributed architecture that co-locates memory and compute, integrating SRAM-based QLoRA adapters to provide massive bandwidth and on-device tunability, and (3) a workload-aware dynamic power gating scheme to minimize energy consumption.
    \item We present a comprehensive evaluation of \aname against general-purpose hardware and other ASICs, demonstrating its state-of-the-art performance and efficiency, thereby establishing a viable path for real-time, high-performance AI on edge devices.
\end{itemize}

\section{Background}
\subsection{Architecture of Large Language Models}


\begin{figure}[htbp]
	\centering
	\includegraphics[width=0.49\textwidth]{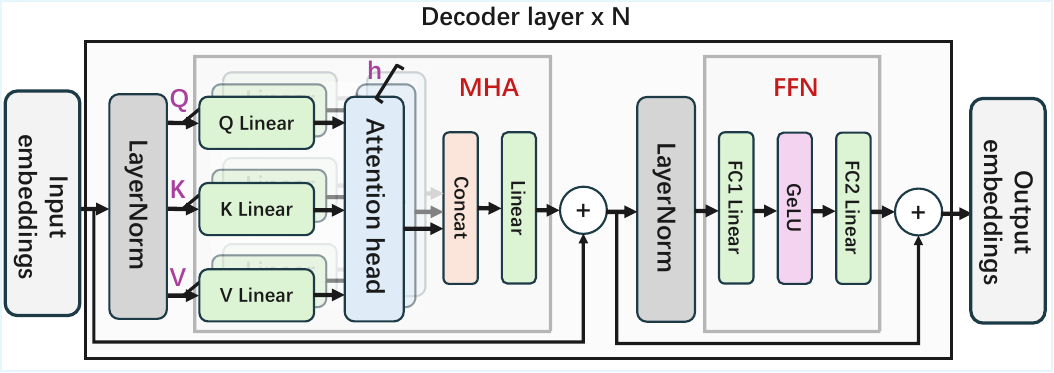}
	\caption{The architecture of a decoder-based Transformer.}
    \label{fig:transformer}
    \vspace{-3mm}
\end{figure}


Current LLMs are based on the decoder-only Transformer \cite{vaswani2017attention}, shown in Fig~\ref{fig:transformer}. The architecture consists of stacked decoder layers, each containing a Multi-Head Attention (MHA) block and a Feed-Forward Network (FFN) block. The MHA block uses query (Q), key (K), and value (V) vectors to compute attention scores, while the FFN block typically uses GeLU \cite{hendrycks2016gaussian} activations. Each block is preceded by LayerNorm \cite{ba2016layer} and utilizes residual connections ~\cite{he2016deep}.

\subsection{Why Ternary LLMs for Edge?}
Deploying LLMs often faces two critical hardware bottlenecks: memory-capacity-bound and memory-bandwidth-bound issues~\cite{li2024large}.
Low-bit LLMs has been a standard method to solve the problem~\cite{zhou2024survey}.
Currently, low-bit LLMs can be produced using two main approaches: (1) quantizing from high-precision LLMs and (2) training low-bit LLMs from scratch. In the first approach, a pre-trained full-precision model is converted to a lower-precision format (e.g., 8-bit or 4-bit) using either post-training quantization (PTQ) or quantization-aware training (QAT)~\cite{zhou2024survey}. PTQ, which requires no retraining, and QAT, which integrates quantization into fine-tuning, have been successfully applied to many LLMs. 
In the second approach, models are trained natively in low precision. A representative example is BitNet b1.58 2B~\cite{bitnet-tech}, a ternary LLM trained from scratch with every weight restricted to three levels (-1, 0, +1). 
Remarkably, BitNet’s ternary weights allow it to match the accuracy of an FP16 model of the same size on standard benchmarks, while reducing memory consumption by 3.55×.

\begin{figure}[htbp]
	\centering
	\includegraphics[width=0.49\textwidth]{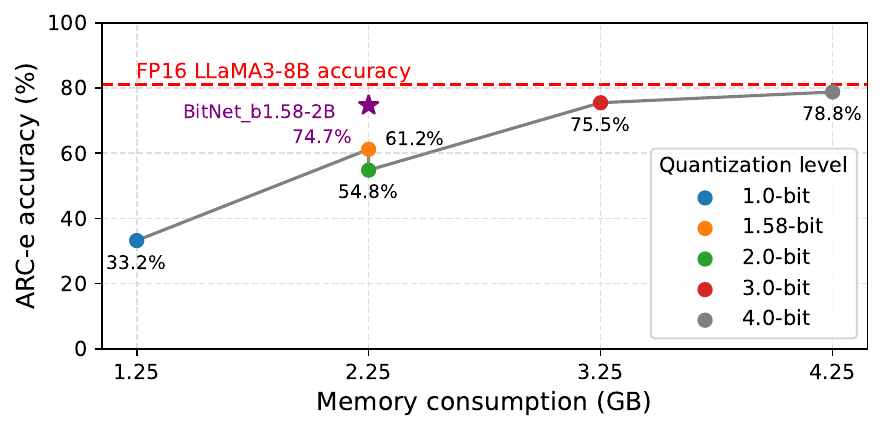}
	\caption{The ARC-e accuracy of LLaMA3-8B with different weight bit-widths~\cite{liu2025paretoq}. The ternary one (1.58-bit) achieves the pareto front considering both accuracy and memory consumption.}
    \label{fig:arc_e_bits}
    \vspace{-3mm}
\end{figure}

Among low-bit models, ternary (1.58-bit) quantization offers a compelling Pareto-front balance between accuracy and memory, as shown in  Fig~\ref{fig:arc_e_bits} for LLaMA3-8B~\cite{dubey2024llama}. Ternary models like BitNet\_b1.58-2B can even match full-precision accuracy\cite{bitnet-tech}, making them ideal for edge deployment. Therefore, designing specialized ternary accelerators is crucial.

\subsection{Deployments of Low-bit LLMs}
Current low-bit LLM deployment efforts span CPUs, GPUs, and ASICs. On CPUs, llama.cpp~\cite{llamacpp} introduces mmap-based integer kernels for edge deployment, while T-MAC~\cite{wei2025t} replaces mixed-precision GEMM with LUT-based lookups, achieving solid inference performance on general-purpose processors. On GPUs, methods like AWQ~\cite{lin2024awq} and Ladder~\cite{wang2024ladder} boost inference efficiency by designing optimized low-bit kernels.
Purpose-built ASICs push performance further. For instance, Guo et al.~\cite{guo2024towards} propose a hybrid PIM architecture based on ReRAM and 3D-stacked SRAM, achieving over 1000 tokens/s. ROMA~\cite{wang2025roma} stores the 2-bit quantized weights of LLaMA3-8B entirely in on-chip ROM, delivering over 20,000 tokens/s at just 33W of power.
Despite the promising performance of these ASIC-based approaches, they often suffer from either insufficient memory capacity or excessive on-chip memory footprint. For example, ROMA integrates 1.86 GB of ROM, 
pushing the total chip area beyond 500 mm\textsuperscript{2} — far too large for edge SoCs (e.g., Apple A17 Pro at 103~mm\textsuperscript{2} ).
Therefore, there is an urgent need to design accelerators that meet the memory capacity and bandwidth requirements of edge LLM inference, while staying within the strict area constraints of edge hardware.

\subsection{Parameter-Efficient Fine-Tuning (PEFT)}

Deploying LLMs on edge devices presents a further challenge when task adaptation or personalization is required. Fully fine-tuning the entire model is computationally prohibitive and memory-intensive, making it infeasible for on-device scenarios. Parameter-Efficient Fine-Tuning (PEFT) methods have emerged as a solution, enabling model adaptation by updating only a small subset of parameters.

Among these techniques, Low-Rank Adaptation (LoRA) \cite{hu2022lora} is a prominent approach. LoRA freezes the large, pre-trained base model weights ($W$) and injects trainable, low-rank matrices ($A$ and $B$) into the network layers. The model's adaptation is then captured in these small adapters ($\Delta W \approx AB$), leaving the original weights untouched. This principle of adapting a model while keeping the base weights immutable makes LoRA a conceptually ideal partner for ROM-based architectures. Furthermore, to minimize the memory and compute overhead of these adapters, subsequent works like QLoRA\cite{qlora} and LoTA-QAF \cite{lora-qaf} have demonstrated the efficacy of quantizing the adapters themselves, for instance, down to ternary weights.
This makes quantized adapters ideal for SRAM storage, where they can be computed sequentially (Base path, then Adapter path) and re-use the same compute units, offering flexibility with manageable overhead \cite{lora-qaf}.
\section{Motivation}

\subsection{Extremely Low Latency Demand when Deploying Edge LLMs}

\begin{figure}[htbp]
	\centering
	\includegraphics[width=0.49\textwidth]{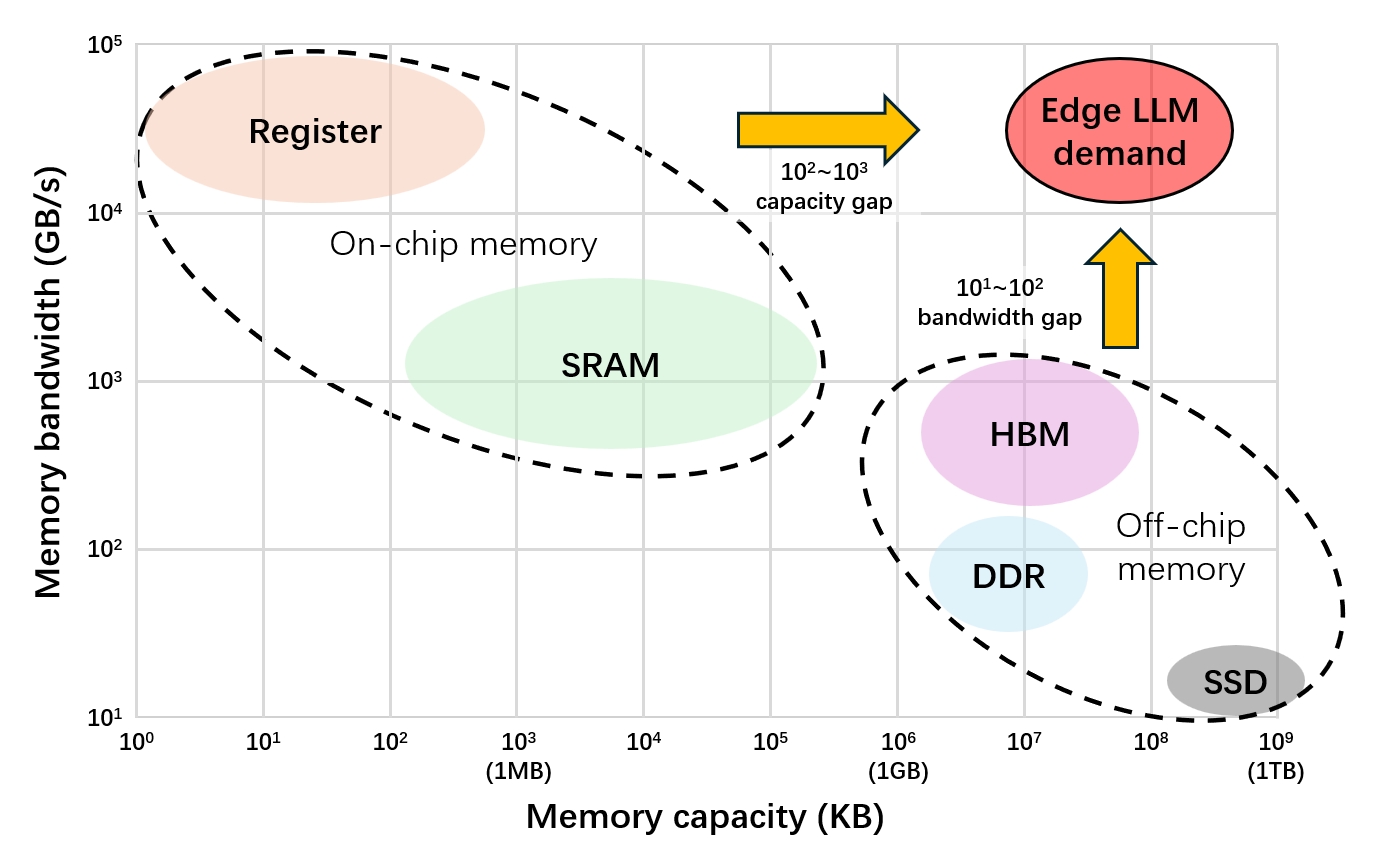}
	\caption{The capacity and bandwidth of different memory. These memory cannot meet either capacity or bandwidth requirements of LLMs on edge.}
    \label{fig:bwbc}
    \vspace{-3mm}
\end{figure}

Currently, certain critical edge scenarios impose extremely stringent latency requirements for LLM inference. In embodied intelligence applications, for instance, generating each action token via an LLM demands a latency below 1 ms, as robots must respond to environmental changes in real-time at frequencies exceeding 1000 Hz~\cite{li2024large}.
Taking the current state-of-the-art robotic arm VLA model, UniVLA~\cite{bu2025univla}, as an example: the LLaMA-2 7B (FP16) model used within it requires 14 GB of storage solely for model weights. During inference, LLaMA-2 generates robotic action tokens in an auto-regressive manner. Since this process is memory-bound, achieving latency under 1 ms per token necessitates a memory bandwidth of at least 14 GB per 1 ms, equivalent to 13.67 TB/s. 
However, existing memory types fail to simultaneously meet the storage and bandwidth needs of these edge scenarios in Fig~\ref{fig:bwbc}: (1) High-bandwidth on-chip memories, such as registers and SRAM, fall short by 2–3 orders of magnitude in memory capacity compared to the requirements of edge LLMs. (2) Off-chip memories with greater storage capacity, such as HBM, have bandwidths that fall short by 1–2 orders of magnitude. Therefore, it is necessary to design an entirely new memory structure tailored for edge LLM inference.

\subsection{ROM to Rescue?}
Read-Only Memory (ROM) is a promising memory structure which has the potential to simultaneously meet the memory capacity and bandwidth demands required for LLM inference. ROM is a type of high-bandwidth on-chip memory, where each memory bit can be accessed using only a single CMOS transistor~\cite{taub1963short}. This approach significantly reduces the area required compared to the conventional 6-transistor (6T) structure used in SRAM. Some ROM-based ASIC designs have demonstrated memory capacities exceeding 1GB, with memory bandwidth reaching hundreds of TB/s. However, despite ROM's substantial improvement in memory density over SRAM, it still falls short of the tens-of-gigabytes storage demands typical of FP16-based LLMs. Fortunately, quantization techniques can greatly reduce the memory requirements of LLMs. For instance, a ternary-quantized 7B LLM, which provides an effective balance between memory demand and model performance, can store all its weights in merely 1.75GB. Hence, combining ROM with ternary LLMs emerges as a highly promising solution for addressing the memory requirements of edge LLM inference.


\begin{figure}[htbp]
	\centering
	\includegraphics[width=0.49\textwidth]{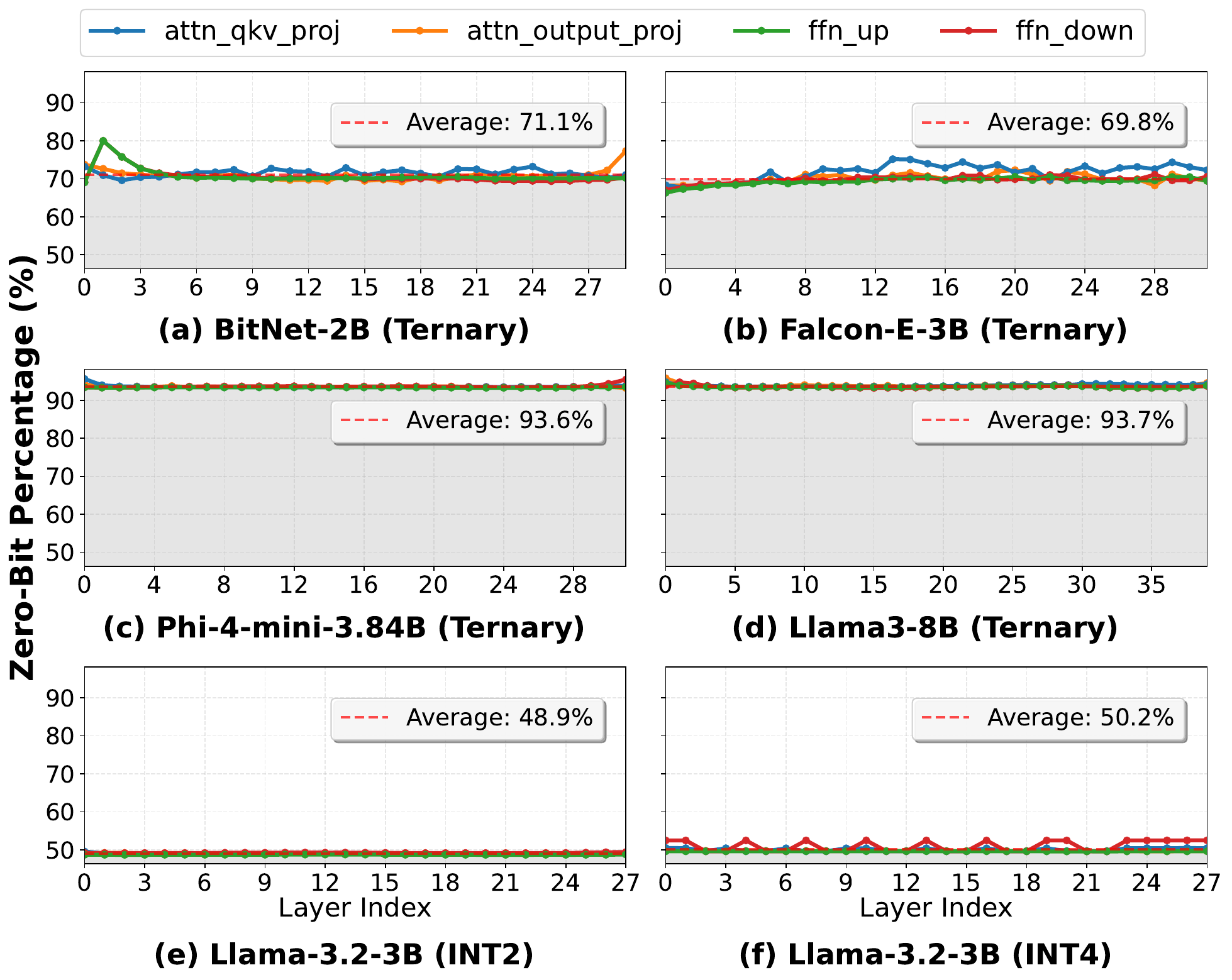}
	\caption{Zero-value bit percentage across ternary model layers. Ternary memory naturally shows high percentage of zero-valued weights. In order to further improve the zero-value bits percentage, we use '10' instead of '11' to encode -1. (a-d) show that the overall sparsity-ratio can be up to 94\%, while INT2 or INT4 quantization typically is around 50\% (e-f). }
    \label{fig:zero_params}
    \vspace{-3mm}
\end{figure}

\subsection{Sparsity Feature of Ternary LLMs can Enhance ROM.}
\label{subsec:ternary weighs motivation}
Although ROM offers substantial memory capacity and bandwidth, existing ROM-based chips are too large in area to be practically deployed at the edge. For example, in ROMA~\cite{wang2025roma}, the on-chip 1.86GB ROM occupies roughly 300 mm², and such a large chip area inevitably leads to increased leakage currents and thus higher power consumption. Fortunately, the inherent sparsity of ternary LLMs provides an opportunity to significantly reduce ROM storage area. Fig~\ref{fig:zero_params} illustrates the sparsity rates (percentage of zero values) in different layers of current ternary LLMs~\cite{ma2024era,almazrouei2023falcon,abdin2024phi,dubey2024llama}. We observe that ternary LLMs trained from scratch, such as BitNet, exhibit sparsity rates exceeding 70\% in most layers, while post-training quantized ternary LLMs achieve sparsity rates as high as 94\% in certain layers. Storing only non-zero weights in ROM could dramatically reduce the required chip area. However, existing sparsity-aware storage methods rely heavily on storing indices of non-zero values, which introduces significant overhead when sparsity rates are moderate or low, ultimately limiting effective capacity reduction. Thus, a new sparsity-aware ternary LLM storage format must be designed. Additionally, current accelerators lack native support for ternary arithmetic, and the overhead introduced by operations like de-quantization further slows down inference speed for low-bit LLMs. Therefore, computational architectures tailored specifically for ternary LLM inference must also be redesigned.
\section{The Architecture of \aname}

\subsection{Architecture Overview}

\aname architecture is designed to overcome the fundamental memory wall that constrains modern large language model inference. The principle of our design is to enable the hardened storage of entire model parameters on-chip, thereby eliminating the costly and power-hungry data transfers to off-chip memory. 
The resulting architecture is a massively parallel, distributed design that pairs storage directly with computation to unlock enormous memory bandwidth.

\begin{figure*}[t]
    \centering
    \includegraphics[width=\textwidth]{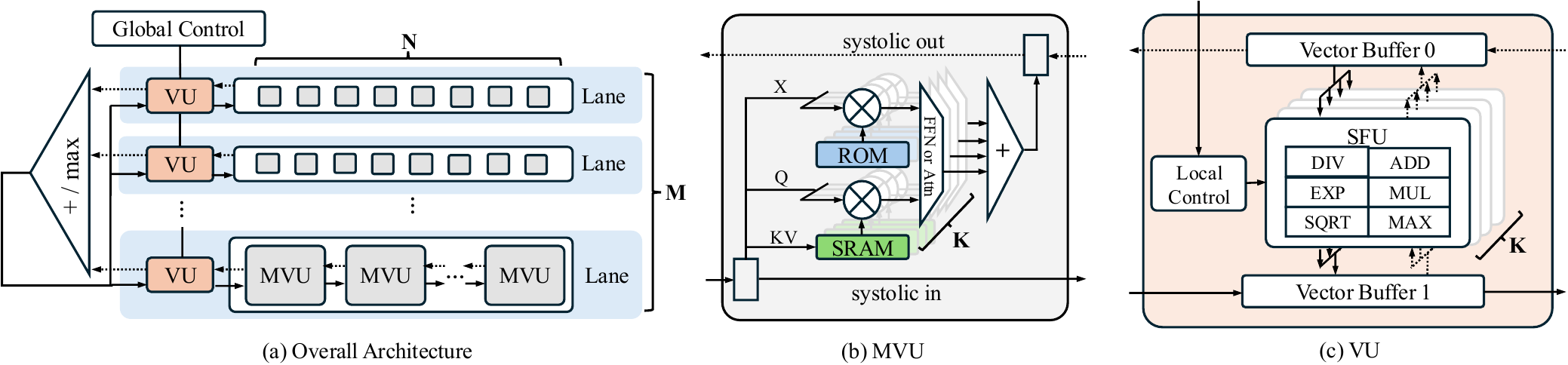}
    \caption{Architecture of \aname. (a) \aname comprise multiple parallel Processing Lanes, each with multiple chained Matrix Vector Units and a shared Vector Unit. (b) Each MVU has its dedicated ROM for weight storage and SRAM for KV cache. The GEMV supports Ternary x FP8 for FFN and FP8 x FP8 for Attention, while sharing the adder tree. (c) Shared Vector Unit in each lane with special arithmetic function. }
    \label{fig:arch-overview}
    \vspace{-3mm}
\end{figure*}

\figref{fig:arch-overview} shows the overview of \aname's architecture. It is composed of a global controller that orchestrates the computation, and several parallel Processing Lanes. 
Each lane contains a shared Vector Unit (VU) and a computation array made of multiple Matrix-Vector Multiplication Unit (MVU). Each MVU contains its own piece of ROM for weight storage and an SRAM for KV Cache storage. 


\subsection{Sparsity-aware ROM for Ternary Weight}

\begin{figure*}[t]
    \centering
    \includegraphics[width=\textwidth]{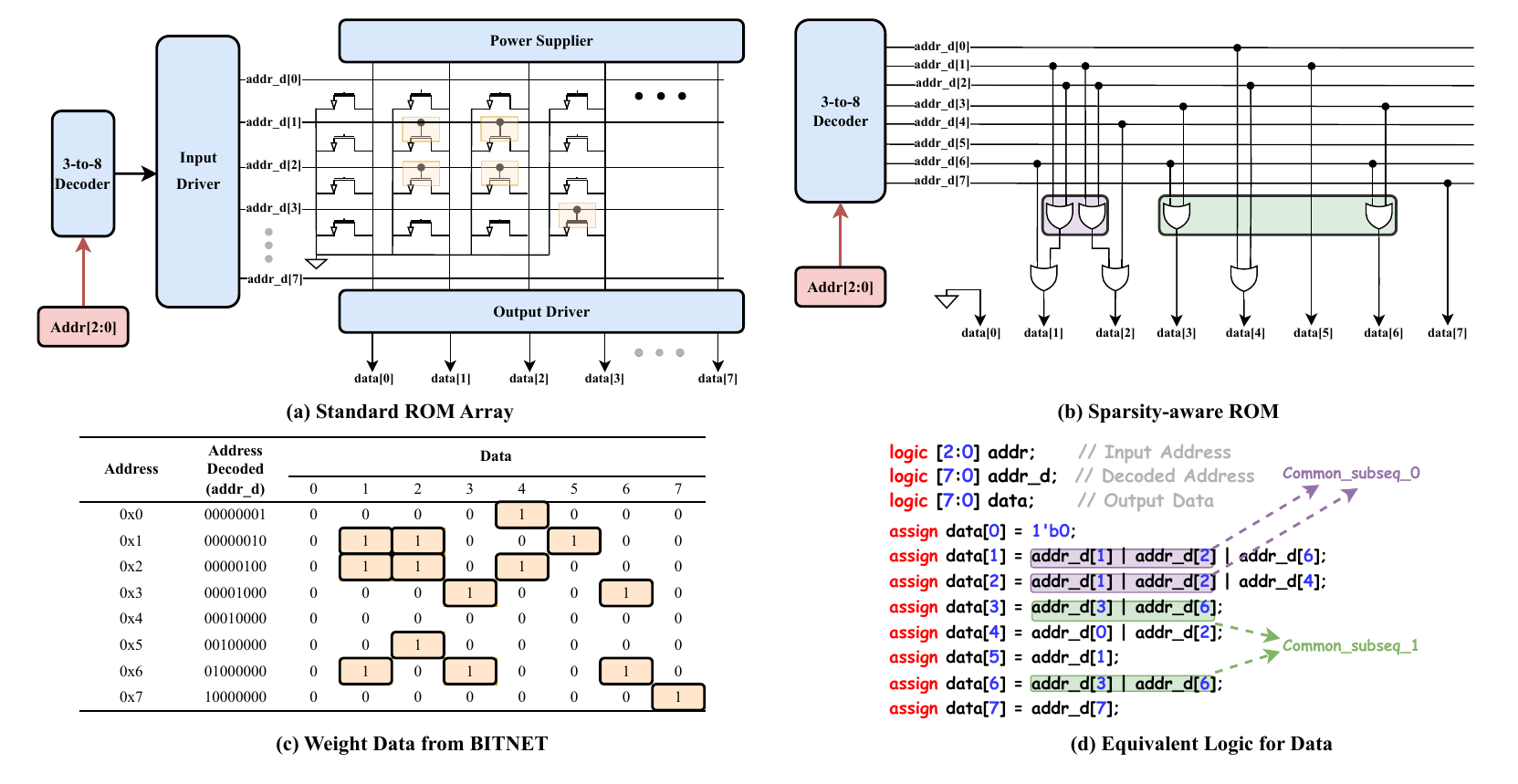}
    \caption{Comparing standard ROM implementation and \aname's sparsity-aware ROM. (a) Standard ROM from Memory Compiler using via-programmed or mask-programmed techniques. (b) Sparsity-aware ROM using combinational logic. (c) Weight data snippet used in the example from a BitNet weight matrix. (d) Using common sub-expression optimization to further reduce area. }
    \label{fig:logic_rom}
    \vspace{-5mm}
\end{figure*}

Storing an entire parameter LLM on-chip requires a leap in memory area density. Existing standard on-chip memory solutions, such as predefined SRAMs and ROMs using memory compiler, are limited by the intrinsic trade-off between storage density and the scalability of the memory array. 
When generating high-density ROMs for weight storage, it results in fixed, rectangular arrays of memory cells where data is physically encoded for each single bits (shown in \figref{fig:logic_rom}(a)) using either via-programmed\cite{via-rom} or mask-programmed\cite{mask-rom} techniques.
This rigid, grid-like layout is suboptimal for ternary models, where a high proportion of zero-bits in weights leads to significant wasted area, as these zero values still consume physical space. Consequently, this standard approach is poorly suited for modern quantized LLMs.

In order to further improve the memory density by leveraging sparsity, we propose a sparsity-aware ROM design by treating the memory content as a simple combinational logic function from input address to one-value bits output rather than a pre-defined physical structure. The output of zero-value bits of input addresses are tied directly ground without generating any logic. \figref{fig:logic_rom}(d) shows the verilog code snippet of the given example.  
This design effectively erases the area cost of all zero-bits of weights. Considering we are targeting on ternary models, the ratio of zero-value bits is extremely high as the zero-valued (represented in '00') parameters are the majority. In addition, the +1 and -1 values are encoded in '01' and '10' in our design to further improve the ratio of zero-value bits. 
This design also leverages common sub-expression elimination in EDA tools (\figref{fig:logic_rom}(c)(d)) to further merge logic and reduce area. In the example, this reduces transistor count from 64 to 28, achieving significant savings.

\subsection{Distributed Processing Architecture}
\label{sec:distributed_arch}

Processing using the on-chip ROM is only efficient if the stored weights can be accessed with sufficient bandwidth. A single, centralized ROM would re-introduce a bottleneck at its access port, undermining the benefits of on-chip storage. Thus, \aname employs a distributed processing architecture that co-locates computation with memory. This strategy not only provides enormous aggregate memory bandwidth by activating all ROM banks in parallel but also minimizes data movement of activation, which is critical for efficiency.

The architecture is organized hierarchically to manage the complexity of LLM inference, consisting of global components that oversee multiple parallel \textit{Processing Lanes}.
As illustrated in Fig~\ref{fig:arch-overview}(a), the top level of the \aname architecture features a set of global components for control and reduction, and a set of parallel \textit{Processing Lanes}.

\paragraph{\textbf{Global Controller and Reduction Tree}}
The global controller orchestrates the overall execution flow by decoding high-level instructions, managing dependencies, and broadcasting control signals to the Processing Lanes.
Alongside the controller, a dedicated \textit{Global Reduction Tree} provides hardware-accelerated reduction operations across all Processing Lanes. This is essential for operations that require aggregating distributed partial results from the parallel lanes, notably for sum (+) and maximum (max) reductions. The aggregate results can be written back to Processing Lanes for subsequent computation.

\paragraph{\textbf{Processing Lanes}}
Each Processing Lane operates in parallel on a distinct portion of the data. Based on the type of layers, e.g. FFN or Attention, the tilling scheme can be slightly different. However, all lanes execute the same instruction stream from the Global Controller but activate a distinct part of the weights stored locally in each lanes in a distributed manner. Direct cross-lane communication is eliminated; lanes only synchronize via the global reduction tree. Consequently, the global reduction tree is the only mechanism for data synchronization and aggregation across the parallel lanes.
Each Processing Lane is a self-contained processing subsystem designed to execute the core operations of a transformer block. It contains two primary components: a \textit{Matrix-Vector Array} composed of multiple chained Matrix-Vector Units (MVU) and a shared \textit{Vector Unit} (VU).

\paragraph{\textbf{Matrix-Vector Units}}
This is the primary engine for performing General Matrix-Vector Multiplications (GEMV).
In our design, each lane consists of an array of MVUs connected in sequence.
While activation data are streamed through the MVUs in a pipelined manner, all partial sums are accumulated locally within each MVU, rather than being propagated downstream.
This structure retains a systolic-like dataflow for activations, but decouples the accumulation path to improve modularity and reduce inter-unit wiring complexity.
An MVU comprises:
\begin{enumerate}
    \item \textbf{ROM for Weights:} Each MVU contains its own dedicated banks of our custom sparsity-aware ROM, which stores a specific sub-matrix of the LLM's weights. This local high-bandwidth allows the architecture to effectively leverage the on-chip bandwidth of the model weights.
    \item \textbf{SRAM for KV Cache:} A small, fast SRAM holds the KV cache required for the local computation. This local storage prevents the need to repeatedly fetch activation data from external memory, thus reducing data movement and power consumption.
    \item \textbf{Compute Unit:} Each MVU contains a compute unit designed to perform the Ternary x FP8 and FP8 x FP8 GEMV operations, which are used in FFN layers and attention layers, respectively. The ternary multiplication as simple as a conditional negation and the adder tree is re-used between the FP8 x FP8 and Ternary x FP8 units to save area. 
\end{enumerate}

\paragraph{\textbf{Vector Unit (VU)}}
While matrix multiplication constitutes the bulk of computation in LLMs, non-linear activation functions, normalization layers, and other element-wise operations are also critical for model accuracy and stability.
To address these, VU is designed as a specialized functional block that efficiently executes vector-wide operations.
It includes a dedicated buffer for storing activations and integrates essential arithmetic logic for computing functions such as exponentiation, division, square root and etc.
This enables the VU to support common operations like Softmax, LayerNorm, and GELU with high throughput and low latency.

\subsection{Operation Scheduling}
The hierarchical and distributed nature of the \aname architecture allows for efficient scheduling of different LLM operators by mapping them to the specialized units best suited for the computation. The scheduling is orchestrated by the Global Controller, which dispatches instructions to the lanes.

\begin{figure}[htbp]
	\centering
	\includegraphics[width=0.49\textwidth]{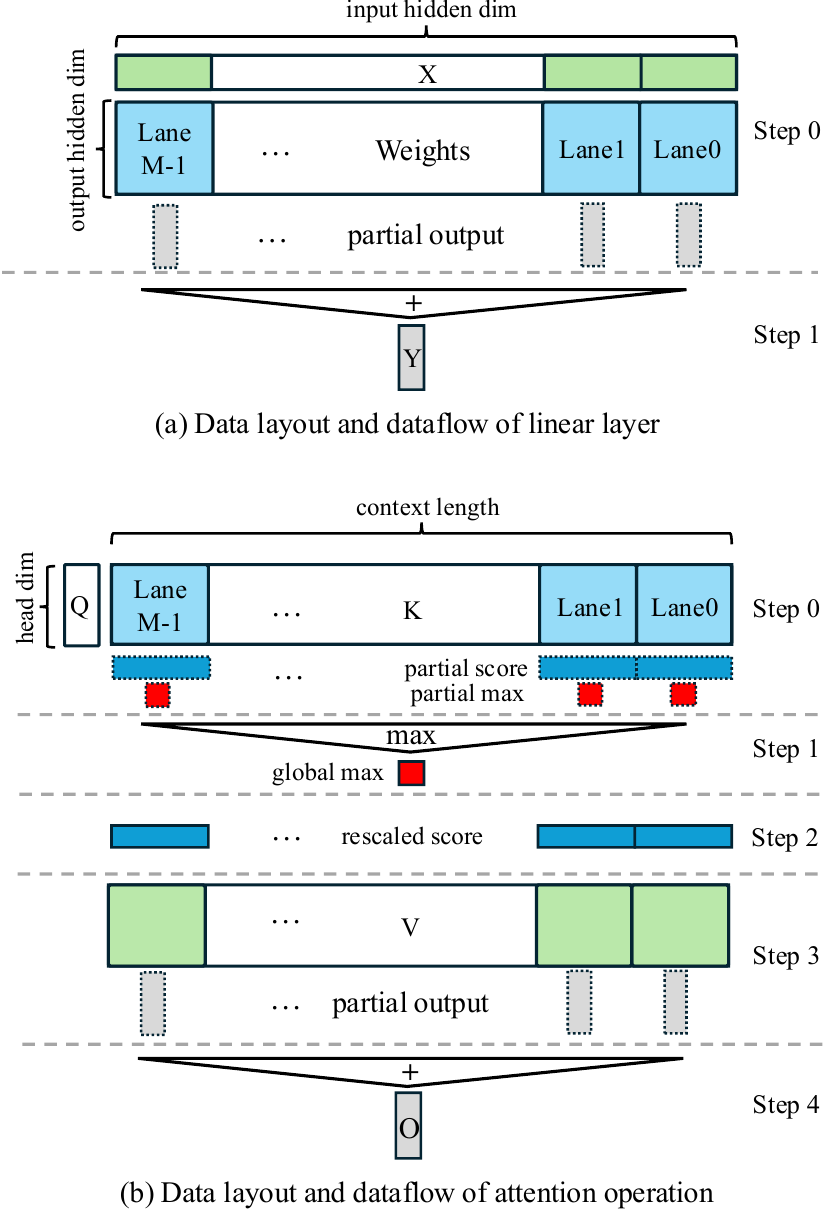}
	\caption{(a) In linear layers, computation are tiled in input hidden dimension of weights in different lanes. The partial results will be aggregate using the global reduction tree. (b) The attention computation is composed of multiple steps. Step 0: compute local attention score and max within the tokens of each lane. Step 1 \& 2: reduce global max and rescale attention score. Step 3 \& 4: compute attention score $\times$ V locally and finally reduce.}
    \label{fig:data_layout}
    \vspace{-5mm}
\end{figure}

\subsubsection{\textbf{Linear Layers}}
Linear layers (FFN, QKVO projection) are mapped as GEMV operations. The FP8 activation vector is tiled across lanes, streamed through the MVUs, and computed against the local ternary weights from ROM. Partial results are aggregated via the global reduction tree.

\subsubsection{\textbf{Attention Operations}}


Our architecture executes all operations token-by-token (similar to decoding), eliminating the prefill/decoding distinction. The KV cache is distributed across the on-chip SRAMs, tiled across the context dimension, similar to flash-decoding~\cite{flash2}.

We adapt the dataflow of flash-decoding with a slight but critical modification to better leverage the unique strengths of our architecture. As shown in \figref{fig:data_layout}(b), it starts from computing the partial attention scores and max within each lane in step 0. However, unlike flash-decoding, which computes a partial, rescaled output locally within each tile, we exploit our powerful global reduction tree. A max reduction operation is performed to find the true global maximum score across all active lanes to rescale the final attention score (step 1 and 2). Finally, the rescaled scores are multiplied with their corresponding V vectors distributively within each lane (Step 3), and a final reduction-sum operation across the lanes produces the complete attention output vector (Step 4). This dataflow is co-designed with our hardware. The original flash-decoding algorithm is optimized to minimize costly off-chip memory accesses and improve the GPU utilization. Since \aname's KV cache is already on-chip, our optimization target shifts to minimizing on-chip computational complexity. By using the fast global reduction tree to establish the global softmax parameters, we bypass the need for intermediate rescaling and accumulation of partial outputs.

\subsubsection{LoRA Operations}

TOM integrates LoRA as its primary flexibility mechanism. Unlike conventional GPU inference where LoRA adapters ($AB$) can be merged with the base model weights ($W' = W + AB$) prior to execution, TOM's base weights in ROM are immutable.

Therefore, TOM adopts a "two-path execution" dataflow:
\begin{itemize}
    \item \textbf{Base Path}: First, the primary computation $h_{base} = W \cdot x$ is executed, with the ternary base model weights $W$ fetched from the high-density ROM banks.

    \item \textbf{Adapter Path}: After the base path is complete, the LoRA computation $h_{lora} = B \cdot (A \cdot x)$ is executed.
\end{itemize}
To implement this efficiently, the LoRA adapter matrices ($A$ and $B$) are stored in on-chip SRAM, sharing resources with the KV Cache. We adopt a quantized adapter approach, using ternary weights for the LoRA matrices. Crucially, this adapter path reuses the existing Ternary $\times$ FP8 compute arrays within our MVUs, minimizing the additional hardware logic required for flexibility. The outputs from the base path ($h_{base}$) and the adapter path ($h_{lora}$) are then summed in the Vector Unit (VU) to produce the final result.

\subsection{\textbf{Workload-Aware Dynamic Power Gating}}


A significant advantage of designing our sparsity-aware ROM from standard logic cells, rather than using traditional compiler-based memory, is the ability to implement fine-grained and aggressive power management. Unlike conventional memory arrays that suffer from long wake-up latencies, our logic-based ROM can be power-gated and restored almost instantaneously without any performance penalty.

\begin{figure}[htbp]
	\centering
	\includegraphics[width=0.49\textwidth]{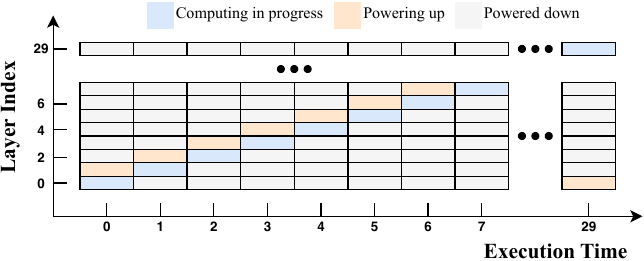}
	\caption{Layer-by-layer dynamic power gating. When executing layer N, layer N + 1's weight will start to power up and the rest will keep powered down.}
    \label{fig:power_flow_pdf}
    \vspace{-3mm}
\end{figure}

We leverage this capability through a workload-aware dynamic power gating strategy. The execution flow of an LLM is predictable, proceeding sequentially from one layer to the next. The Global Controller, which orchestrates this flow, is aware of which layer is currently active. Consequently, it only enables power to the specific ROM banks that store the weights for the currently executing layer. All other ROM blocks associated with inactive layers are completely power-gated, eliminating their static power consumption.

As illustrated in \figref{fig:power_flow_pdf}, the set of active ROM banks changes dynamically as the inference process moves through the different layers of the neural network. This simple yet highly effective technique drastically reduces the overall static power consumption of the chip, as the majority of the on-chip weight memory is powered down at any given moment. This is achieved with zero impact on latency, making it a pure efficiency gain.

\section{Evaluation}

This section presents a comprehensive evaluation of \aname architecture. We begin by detailing our experimental setup, including the system configuration, baseline platforms, and evaluation workloads. Subsequently, we conduct an in-depth analysis of our core innovation—the area density of the Sparsity-Aware ROM. Finally, we present end-to-end system performance and power efficiency results, comparing \aname with state-of-the-art designs to demonstrate its overall architectural advantages.


\subsection{Evaluation Setup}

\paragraph{\textbf{System Configuration}}

The evaluation was performed on the TOM architecture detailed in Section IV. The specific hardware configuration used for these tests is detailed in Table~\ref{tab:hardware_config}. 
\begin{table}[]
\centering
\caption{\aname Architectural Configuration}
\label{tab:hardware_config}
\renewcommand{\arraystretch}{1.3}
\begin{tabular}{lcc}
\hline
\textbf{Component}                        & \textbf{Parameter}    & \textbf{Value} \\ \hline
\multirow{3}{*}{\textbf{Top-Level}}       & Frequency             & 500 MHz             \\
                                          & Processing Lanes (M)  & 16             \\
                                          & Total on-chip ROM     & 498.54 MB       \\
                                          & Total on-chip SRAM    & 37. 5 MB       \\ \hline
\multirow{2}{*}{\textbf{Processing Lane}} & MVUs per Lane (N)     & 10             \\
                                          & Vector Unit Width (K) & 16             \\ \hline
\multirow{2}{*}{\textbf{MVU}}             & Weight Capacity       & 3180 KB         \\
                                          & KV Cache Capacity     & 240 KB        \\ \hline
\multirow{3}{*}{\textbf{Others}}          & Weight Format         & Ternary        \\
                                          & Act./KV Cache Format  & FP8            \\
                                          & Max on-chip Context   & 1024           \\ \hline
\end{tabular}
\vspace{-5mm}
\end{table}

\paragraph{\textbf{Baselines}}

We evaluate \aname against two categories of baselines: general-purpose processors for direct performance comparison and domain specific hardware designs for conceptual and architectural comparison.

\begin{itemize}
    \item \textbf{CPU:} Intel Core i5-12500H processor running bitnet.cpp\cite{bitnetcpp}, a highly optimized framework for BitNet-like models, representing the performance of mainstream edge devices.
    \item \textbf{GPU:} NVIDIA A100 (40G) datacenter GPU, also running a GPU-optimized version of bitnet.cpp~\cite{bitnetcpp}, to benchmark against high-performance general-purpose architectures. Since TOM is designed for low-latency, single-stream edge scenarios, all GPU evaluations are conducted with a batch size of 1 to ensure a fair, application-aligned comparison.
    \item \textbf{ASICs/PIMs:} For architectural comparison on key metrics like energy efficiency, we benchmark TOM against several published ASIC and Processing-in-Memory (PIM) designs. The specific designs included in our comparison are Olive\cite{olive}, FIGNA\cite{figma}, Spatten\cite{spatten}, TF-MVP\cite{tf-mvp}, Arc\cite{arc}, and SOFA\cite{sofa}.
\end{itemize}


\begin{figure}[htbp]
	\centering
	\includegraphics[width=0.49\textwidth]{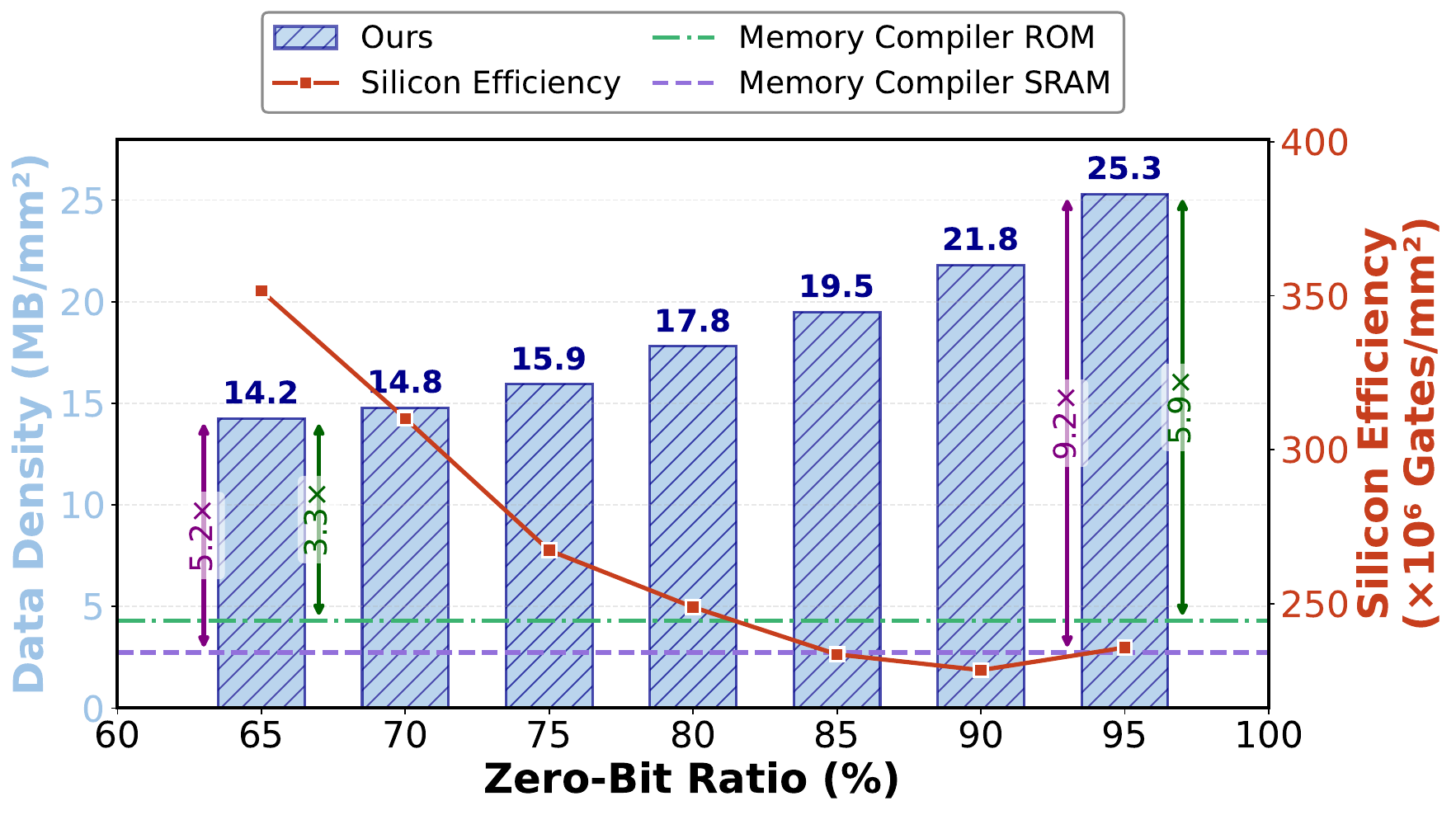}
	\caption{The Effect of Sparsity on ROM Area Density. This chart plots the storage density (left Y-axis) and silicon efficiency (right Y-axis) of a 2048x128 sparsity-aware ROM as a function of the weight's zero-bit ratio.}
    \label{fig:density_vs_sparsity}
    \vspace{-3mm}
\end{figure}

\paragraph{\textbf{Hardware Implementation and Workload}}

Our design is implemented in Verilog HDL and evaluated using an industry-standard Electronic Design Automation (EDA) toolchain. The SFU implements the non-linear operations (square, division, and exponentiation) using reusable, silicon-proven components from the Synopsys DesignWare (DW) Foundation Library.

The subsequent design flow included logic synthesis using Synopsys Fusion Compiler to map the Verilog code to a target technology (for area and timing results), followed by detailed power analysis using PowerArtist.

To validate our synthesis-based results, we performed a full Place \& Route (P\&R) analysis on the core sparsity-aware ROM modules, which represent the majority of the chip's area. The P\&R results showed <5\% deviation in area and power from our synthesis estimates. This high fidelity confirms the reliability of our full-chip analysis, owing to the regular, logic-based structure of our ROM which minimizes routing overhead.

We measure the overall system performance via cycle-accurate, high-speed simulation using Verilator. We primarily evaluate the performance on the BitNet-2B\cite{bitnet-tech} model, using tasks with varying input and output sequence lengths (e.g., 64, 128, 256, and 512 tokens) to comprehensively measure performance.

\begin{figure}[htbp]
	\centering
	\includegraphics[width=0.49\textwidth]{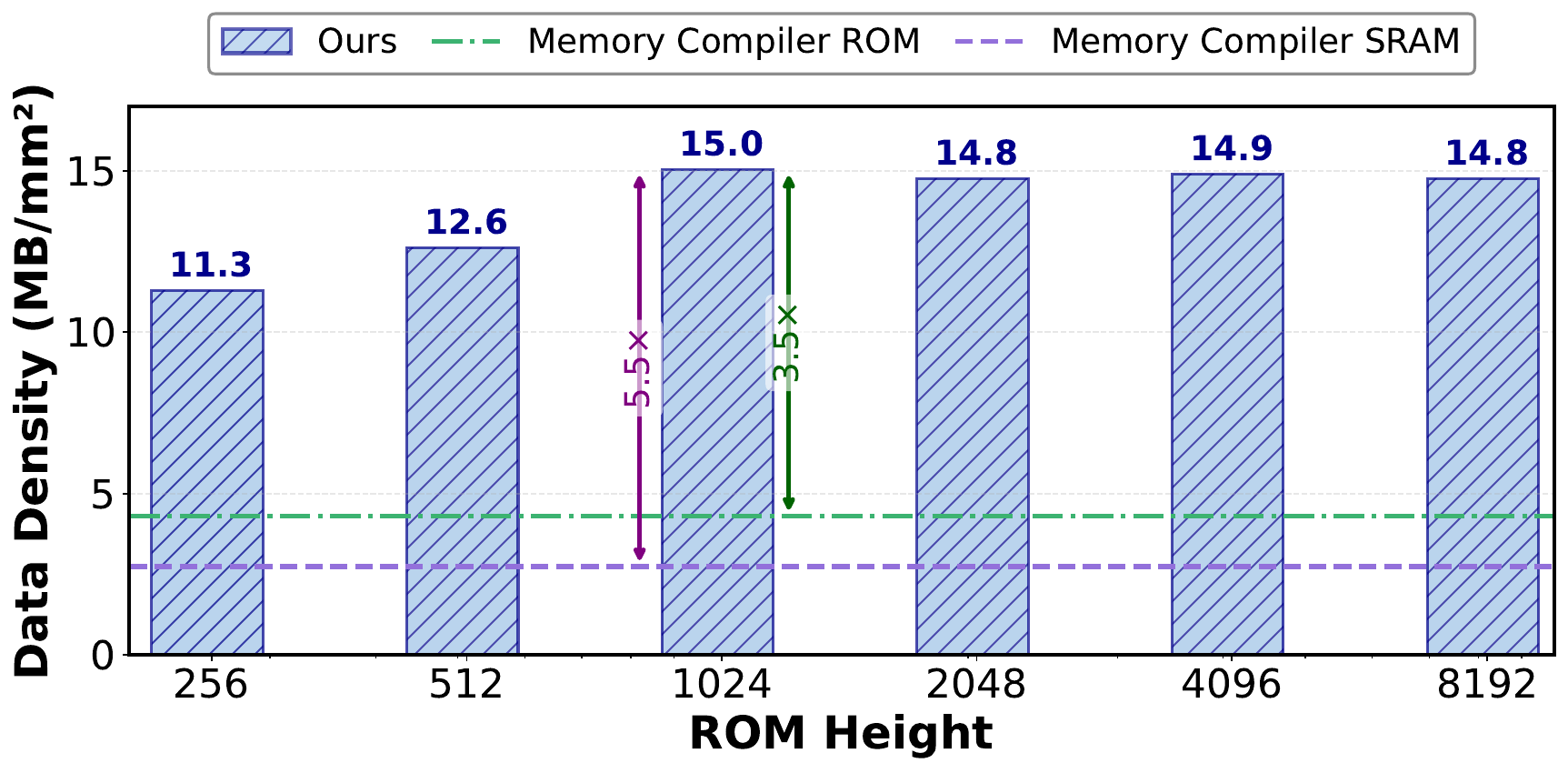}
	\caption{The Impact of ROM Bank Height on Area Density. This chart plots the area density (Y-axis) versus the ROM bank height (X-axis). For this analysis, the ROM bank width is fixed at 128, and the weight zero-bit ratio is fixed at 70\% to find the optimal bank geometry for synthesis.}
    \label{fig:density_vs_bank}
    \vspace{-3mm}
\end{figure}

\subsection{Area Density of Sparsity-Aware ROM}

\paragraph{\textbf{Impact of Sparsity on Area Density}}

Our experimental results, presented in Fig~\ref{fig:density_vs_sparsity} for a 2048x128 ROM bank, confirm a strong positive correlation between the zero-bit ratio in the weights and the achievable storage density of our logic-based ROM. For instance, at a 65\% zero-bit ratio, our ROM achieves a density of 14.2 MB/mm\textsuperscript{2}, which increases to an impressive 25.3 MB/mm\textsuperscript{2} as the ratio reaches 95\%.

We compare \aname's memory density against prior academic and industrial designs to assess its relative standing. All results are normalized to the 7nm node for a fair comparison. We use scaling factors derived from compiler-generated ROMs across different nodes (Table~\ref{tab:memory_density}), yielding a 12.04x scaling factor from 65nm to 7nm and a 3.28x factor from 28nm to 7nm. 

\begin{table}[h]
    \centering
    \caption{Reference Area Densities of a Compiler-Generated ROM Across Multiple Process Nodes. All ROMs have a fixed size of 2048x64.}
    \label{tab:memory_density}
    \setlength{\tabcolsep}{12pt} 
    \renewcommand{\arraystretch}{1.2} 
    \begin{tabular}{ccc}
        \toprule
        \textbf{Process (nm)} & \textbf{Area Density (MB/mm\textsuperscript{2})} & \textbf{Ratio} \\
        \midrule
        65                    & 0.357                           & 1.00           \\
        28                    & 1.308                           & 3.66           \\
        7                     & 4.30                            & 12.04          \\
        \bottomrule
    \end{tabular}
\end{table}






Fig~\ref{fig:density_vs_sparsity} also plots the silicon efficiency, measured in synthesized gates per mm\textsuperscript{2}, which reveals a subtle design trade-off. While higher sparsity reduces the absolute number of logic cells required, at extreme levels of sparsity, the increased routing complexity from irregular logic placement can slightly diminish the area gains per gate. Nevertheless, even with this second-order effect, the overall density advantage remains substantial across the entire realistic sparsity range for ternary LLMs. Compared to memories generated by a standard memory compiler on the same technology node, \aname's ROM density is 5.2x higher than a standard ROM and 3.3x higher than a standard SRAM at a 65\% zero-bit ratio.



We also analyzed ROM bank granularity (Fig~\ref{fig:density_vs_bank}), fixing sparsity at 70\% and width at 128. Data density peaks at 15.0 MB/mm\textsuperscript{2} at a height of 1024. This non-monotonic behavior reflects the synthesis tool's optimization (e.g., common sub-expression elimination), which finds a 'sweet spot' at 1024-height, balancing optimization scope against routing complexity.

\begin{table}[]
\centering
\caption{Memory Density Comparison Across Different Works and Technologies. For technologies other than 7nm, the density is also scaled to 7nm for comparison.}
\renewcommand{\arraystretch}{1.3}
\label{tab:memory_density_comparison}
\begin{tabular}{ccccc}
\toprule
\multirow{2}{*}{\textbf{Method}} &
  \multirow{2}{*}{\textbf{\begin{tabular}[c]{@{}c@{}}Tech\\ (nm)\end{tabular}}} &
  \multirow{2}{*}{\textbf{Device Type}} &
  \multicolumn{2}{c}{\textbf{Density (MB/mm\textsuperscript{2})}} \\ \cline{4-5} 
                             &    &             & \textbf{@Tech} & \textbf{scaled @7nm} \\ \hline
ISSCC'24\cite{isscc-24}      & 7  & 3D-SRAM     & \textbf{4.0}            & -                    \\
MICRO'22\cite{micro-22}      & 7  & 3D-DRAM     & \textbf{8.4}            & -                    \\
CICC'24\cite{cicc-24}        & 28 & MLC-ROM     & 1.09           & \textbf{3.57}                 \\
ASSCC'24\cite{asscc24-rom}   & 28 & QLC-ROM     & 2.46           & \textbf{8.06}                 \\
ASPDAC'25\cite{aspdac25-rom} & 65 & Digital ROM & 0.06           & \textbf{0.72}                 \\
\textbf{\aname}                       & 7  & Digital ROM & \textbf{15.0}          & -                    \\ \bottomrule
\vspace{-6mm}
\end{tabular}
\end{table}


\paragraph{\textbf{Comparison with State-of-the-Art}}



\begin{table}[htbp]
    \centering
    \caption{Aggregate Memory Bandwidth and Capacity of TOM Versus High-Performance Architectures.}
    \begin{tabular}{lcc}
        \toprule
        \textbf{Design}       & \textbf{Bandwidth}   & \textbf{Total Capacity} \\
        \midrule
        3D SRAM\cite{isscc-24}              & 64 GB/s per Unit         &    16 MB \\
        3D DRAM\cite{micro-22}              & 16 GB/s per Unit         &    32 MB \\
        H100 (HBM3e)\cite{nvidia2024gracehopper} & 4.8 TB/s per GPU    &    144 GB \\
        Cerebras (SRAM)\cite{cerebras}      & 255 TB/s per Die         &    44 GB \\
        \textbf{\aname}                  & \textbf{200 TB/s}        &  \textbf{ 536.04 MB}  \\
        \bottomrule
    \end{tabular}
    \label{tab:bandwidth_comparison}
\end{table}

\begin{figure}[htbp]
\centering	\includegraphics[width=0.49\textwidth]{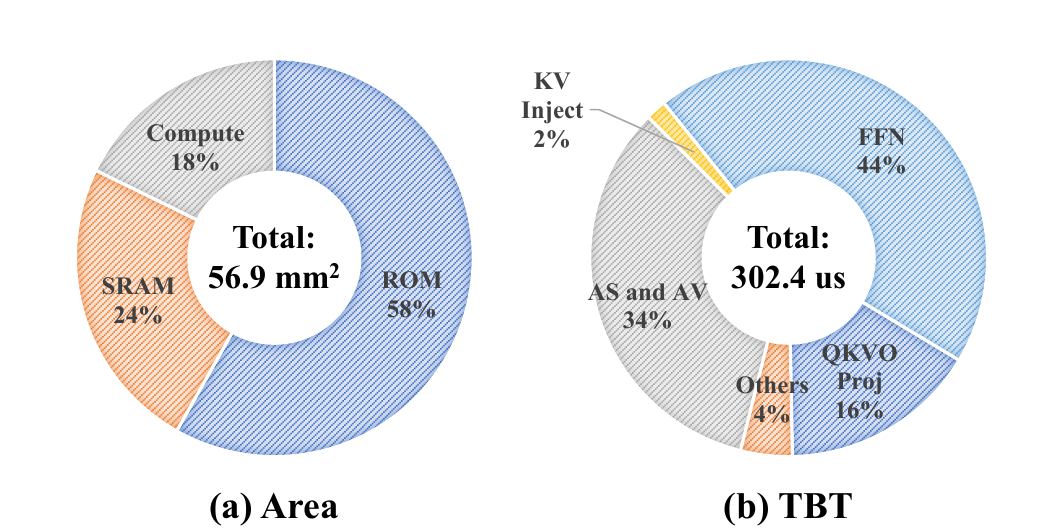}
	\caption{Area and Latency Breakdown of the TOM Architecture. (a) Area Decomposition. (b) TBT Breakdown, showing the latency decomposition for generating a single token.}
    \label{fig:decomposition}
    \vspace{-5mm}
\end{figure}

\phantomsection\label{B-Q1}
For a model like BitNet, which has approximately 40\% zero-valued weights, the corresponding zero-bit ratio is about 70\%. At this level of sparsity, \aname achieves a storage density of 15.0 MB/mm\textsuperscript{2}. 

To evaluate this result, we compare it with the state-of-the-art in high-density on-chip memory, which includes several noteworthy ASIC designs. These benchmarks range from fully digital, sparsity-aware compute-in-ROM like DCIROM~\cite{aspdac25-rom}, to analog hybrid SRAM/MLC-ROM CIM~\cite{cicc-24}, and non-sparsity-aware designs using advanced 4-bit/transistor Quad-Level Cell (QLC) ROM~\cite{asscc24-rom}.

As detailed in Table\ref{tab:memory_density_comparison}, this figure not only surpasses other digital ROM designs but also exceeds complex 3D-stacked solutions. 

Specifically, our on-chip 2D solution is approximately 75\% denser than 3D-stacked DRAM \cite{micro-22} (8.4 MB/mm²). This demonstrates that our sparsity-aware approach achieves exceptional efficiency without resorting to costly and complex 3D integration.

\begin{figure}[htbp]
	\centering	\includegraphics[width=0.3\textwidth]{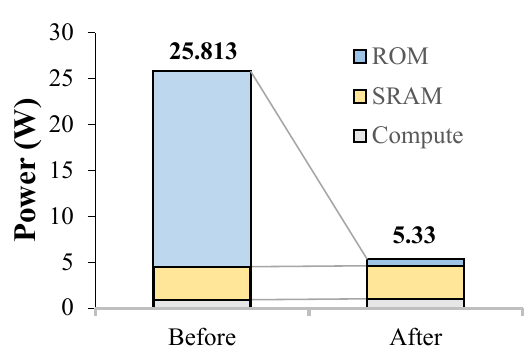}
	\caption{Power Breakdown Before and After Applying Workload-Aware Dynamic Power Gating.}
    \label{fig:power_gating}
    \vspace{-5mm}
\end{figure}

\begin{figure*}[t]
    \centering
    \includegraphics[width=\textwidth]{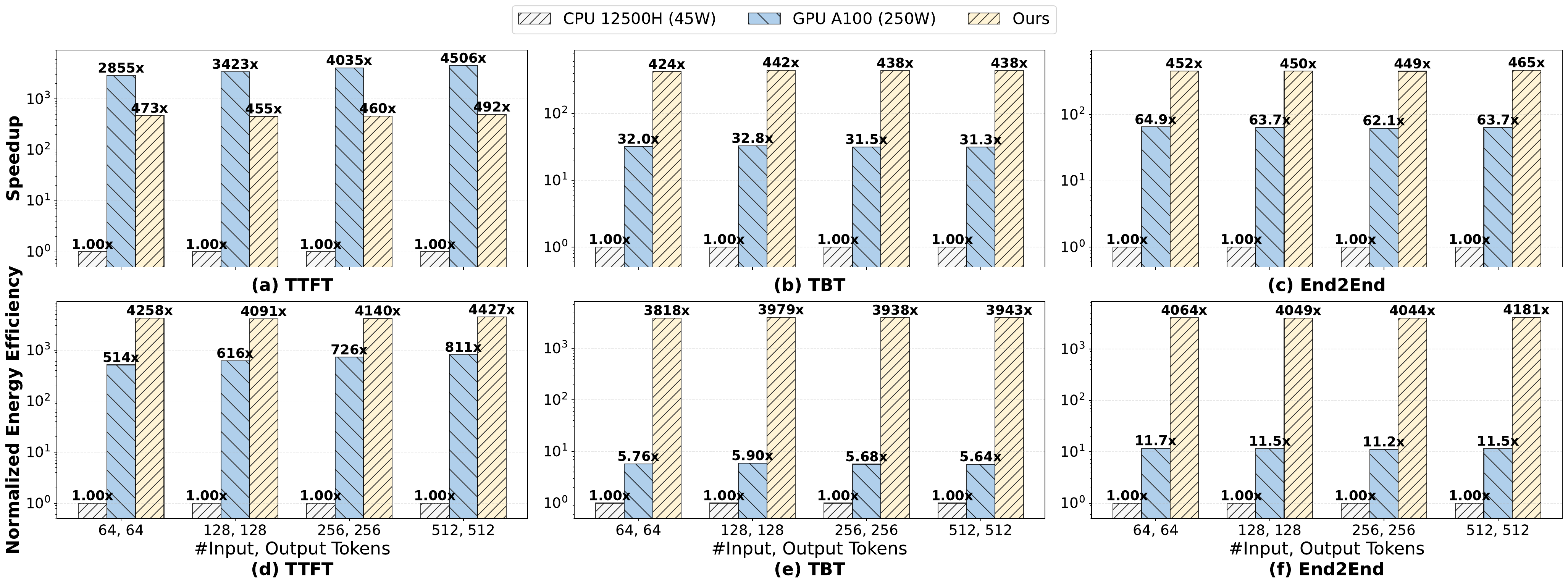}
    \caption{Performance and Energy Efficiency Comparison Across Platforms. These charts compare the performance of \aname against a CPU (Intel i5-12500H) and a GPU (NVIDIA A100) on various workloads. (a-c) Plot the speedup for TTFT, TBT, and End-to-End latency. (d-f) Plot the normalized energy efficiency for the same metrics. Energy efficiency here is defined as performance per Watt.}
    \label{fig:arch-latency_comparison}
    \vspace{-5mm}
\end{figure*}

\subsection{Performance Analysis}

\paragraph{\textbf{Latency and Throughput Analysis}}



As shown in Table~\ref{tab:bandwidth_comparison}, TOM achieves an aggregate on-chip bandwidth of 200 TB/s, which is over 41x greater than an NVIDIA H100 GPU\cite{nvidia2024gracehopper}. This massive bandwidth directly translates to a significant reduction in token generation latency.

While a wafer-scale design like Cerebras\cite{cerebras} achieves a higher peak bandwidth (255 TB/s), it does so with a much larger area and power budget, making it unsuitable for the edge. \aname, in contrast, provides this substantial bandwidth within a practical chip footprint. 

\paragraph{\textbf{Overall Performance}}

Our architecture evaluates both the prefill phase, measured as Time to First Token (TTFT), and the subsequent token-by-token decoding phase (Time Between Tokens, TBT).

Consequently, in terms of overall end-to-end performance, \aname achieves a commanding speedup of up to 63.7x over the A100. This results in a peak throughput of approximately 3306 TPS, a figure that not only enables fluid, real-time human-computer interaction but also meets the stringent sub-millisecond latency requirements of future edge applications like robotics and augmented reality.

A detailed breakdown of the latency for generating a single token is provided in Fig~\ref{fig:decomposition} (b). The 302.4 µs latency per token enables a peak throughput of 3306 TPS. The analysis reveals that the FFN layers are the most time-consuming component, accounting for 44\% of the latency. The core attention mechanism, including score calculation and value multiplication (AS and AV), constitutes 34\% of the time. 

\subsection{Power and Energy Efficiency Analysis}

The overall area and latency breakdowns of the TOM architecture are detailed in Fig~\ref{fig:decomposition}. The chip's total area is 56.9 mm² , with the majority (58\%) allocated to our high-density, sparsity-aware ROM. 
The on-chip SRAM for the KV cache and the distributed compute logic account for 24\% and 18\% of the area, respectively. 

\paragraph{\textbf{Workload-Aware Dynamic Power Gating}}

To quantify this benefit, we conducted an ablation study. As shown in Fig~\ref{fig:power_gating}, with dynamic power gating disabled, the total chip power is 25.813W, the majority of which (21.306 W) comes from the power of all ROM banks. After enabling the workload-aware dynamic power gating technique, the total power drops dramatically to 5.33W—a reduction of nearly 80\%. 

\begin{figure}[htbp]
	\centering
	\includegraphics[width=0.49\textwidth]{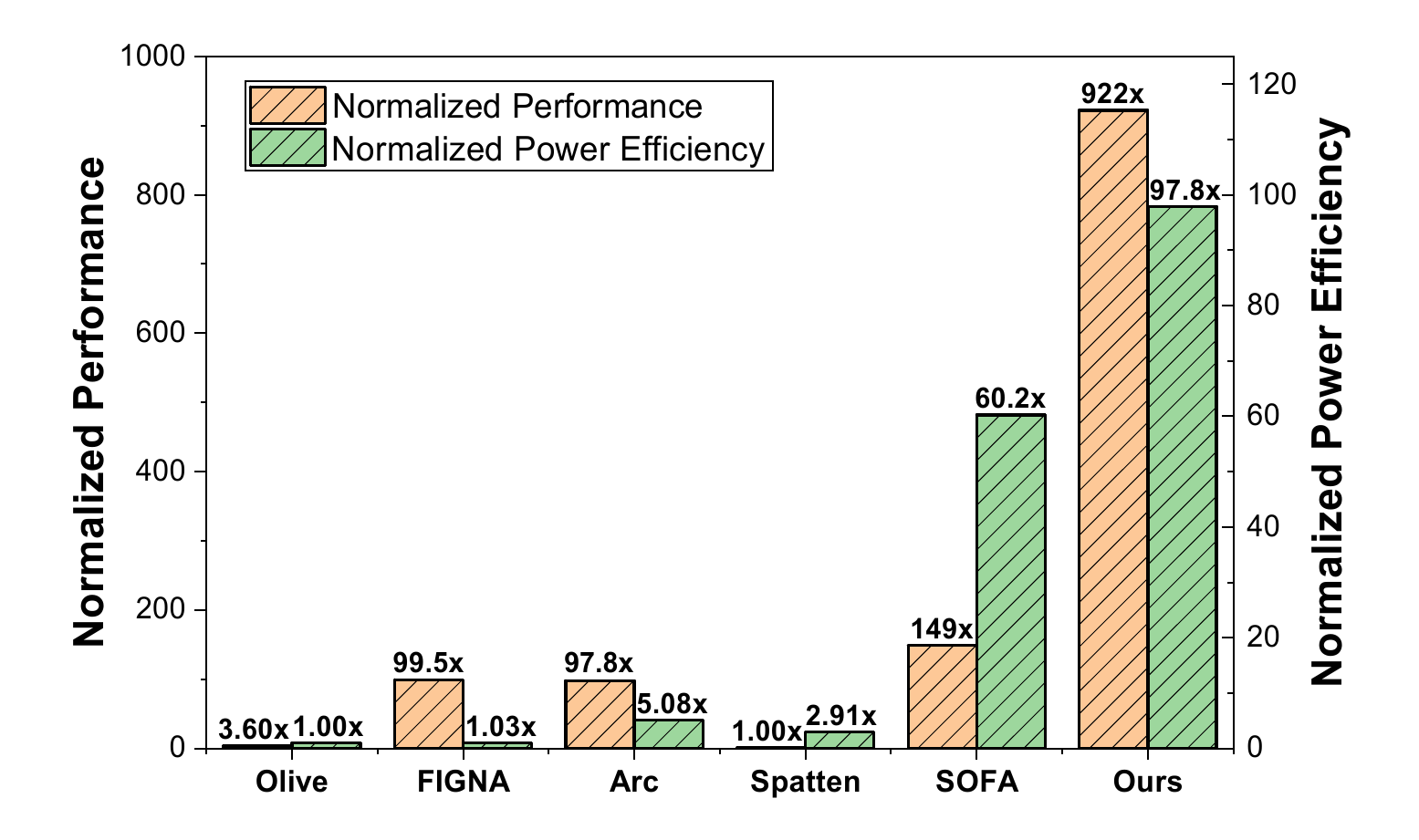}
    \caption{Normalized Performance and Power Efficiency Comparison with State-of-the-Art ASIC and PIM Designs. Performance is normalized to Spatten \cite{spatten}, and Power efficiency (TOPS/W) is normalized to Olive \cite{olive}.}
    \label{fig:efficiency_comparison}
    \vspace{-4mm}
\end{figure}

\paragraph{\textbf{Efficiency Comparison}}

The architectural principles that drive \aname's performance also yield exceptional energy efficiency. To create a fair and comprehensive comparison across a diverse range of platforms, we distinguish between two metrics. For general-purpose processors like CPUs and GPUs, we use energy efficiency, which we define as the application-level performance (TTFT, TBT, and End-to-End throughput) per unit of power, reflecting the real-world speed a user would experience. However, for comparisons against other specialized ASICs and PIMs, a direct comparison of application throughput is often challenging, as these designs frequently employ different models and unique hardware configurations. Therefore, using power efficiency—defined as the computational throughput (TOPS) per Watt—provides a more stable and equitable baseline for assessing the fundamental efficiency of the hardware itself. This dual-metric approach ensures a robust evaluation of \aname's advantages.

First, when compared to general-purpose processors, \aname's advantage in energy efficiency is overwhelming. As shown in Fig~\ref{fig:arch-latency_comparison}(d)-(f), TOM's end-to-end normalized energy efficiency is over 60x higher than the NVIDIA A100 GPU (e.g., 63.7x for the 256/256 token task) and over 4000x higher than the CPU. 

Second, we compare TOM against other state-of-the-art ASIC and PIM designs in Fig~\ref{fig:efficiency_comparison}, evaluating both normalized performance and normalized power efficiency. To create a fair performance baseline across different models, quantizations, and hardware, performance is normalized using available TPS, operating frequency, and model size (relative to Spatten \cite{spatten}. Power efficiency (TOPS/W) is normalized to Olive \cite{olive}.This comparison highlights the comprehensive advantages of our architecture. On the performance front, TOM achieves a 922x speedup, vastly outperforming SOFA \cite{sofa} (149x) and Arc \cite{arc} (97.8x). In terms of power efficiency (TOPS/W), TOM demonstrates a 97.8x improvement over the Olive baseline, similarly outperforming SOFA (60.2x) and Arc (5.08x). This dual superiority in both speed and efficiency validates our co-design approach.


\begin{figure}[htbp]
	\centering
	\includegraphics[width=0.49\textwidth]{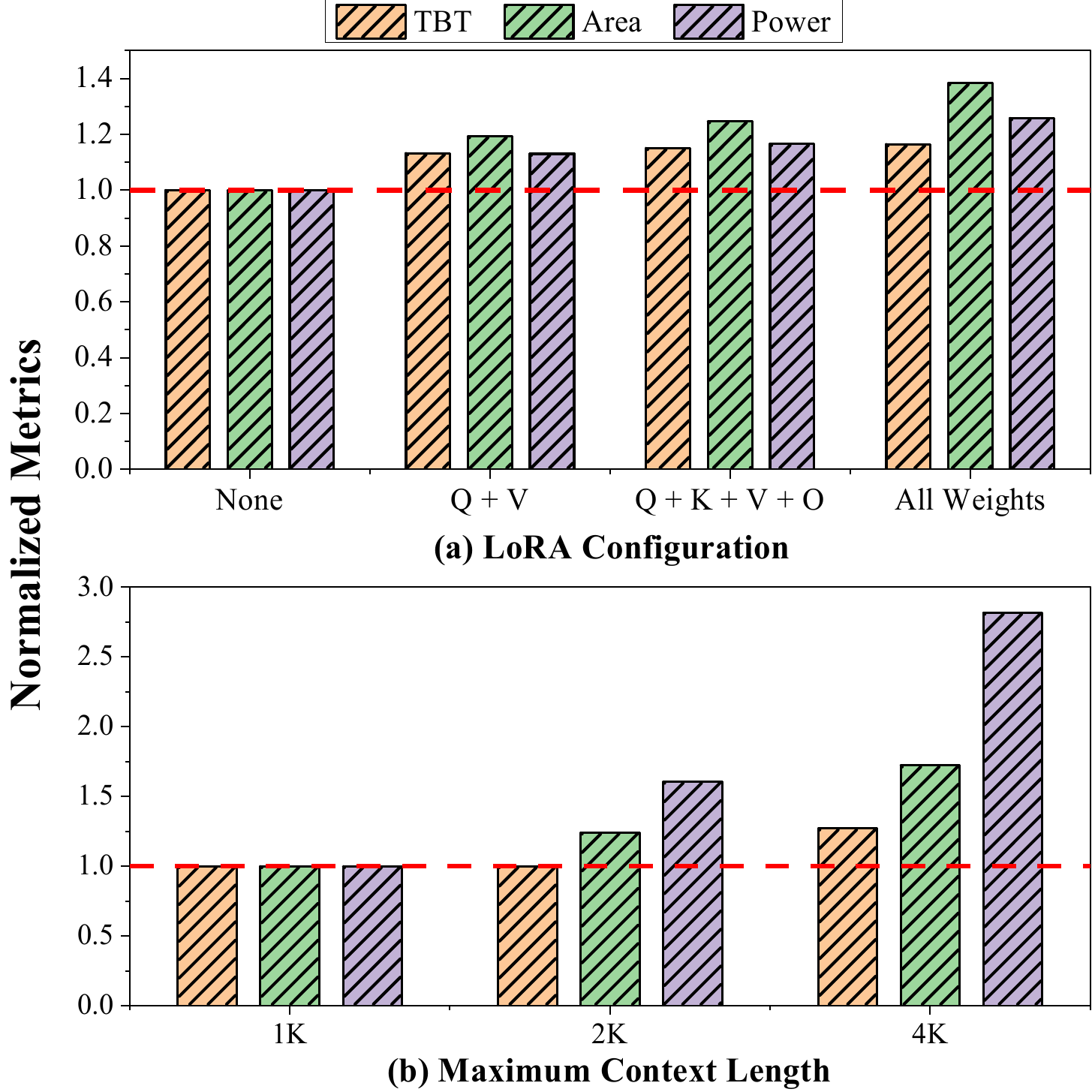}
	\caption{Normalized Overhead for LoRA and Context Scaling. (a) plots the impact of applying LoRA to various weight matrices. (b) plots the scaling impact of increasing the maximum context length.}
    \label{fig:scaling}
    \vspace{-5mm}
\end{figure}

\phantomsection\label{CC-Q2}
\subsection{Scaling Analysis}


\paragraph{\textbf{Scaling with LoRA}}

We analyzed the overhead of integrating LoRA for on-device fine-tuning. We adopt an approach similar to LoTA-QAF \cite{lora-qaf}, storing ternary LoRA matrices in on-chip SRAM, which is shared with the KV Cache. Figure~\ref{fig:scaling}(a) shows the normalized overhead on TBT, area, and power for various common configurations (None, Q+V, Q+K+V+O, All Weights)\cite{hu2022lora}. The results demonstrate that the cost of applying LoRA is acceptable, primarily impacting SRAM area and power with a modest increase in TBT, which we consider a worthwhile trade-off for the flexibility gained.

\paragraph{\textbf{Scaling with Context Length}}
We analyzed the impact of increasing the maximum context length on TBT (Time-Between-Tokens), area, and power, as shown in Figure~\ref{fig:scaling}(b). The TOM architecture possesses inherent computational redundancy when processing Attention. Consequently, TBT does not increase significantly as the context length scales (e.g., up to 2560 tokens). The primary cost of supporting longer contexts is confined to the on-chip SRAM required for the KV Cache. The area and power of this SRAM scale linearly and predictably with the context length, which is an unavoidable trade-off.

\section{Related Works}

\subsection{ASIC-based Acceleration}


ASIC accelerators co-design hardware for low-bit quantization by handling outliers \cite{olive}, avoiding dequantization \cite{figma}, decomposing matrices \cite{tender}, or unifying quantization \cite{m-ant}. Other custom accelerators exploit various forms of sparsity by pruning tokens and heads \cite{spatten}, predicting attention sparsity \cite{sofa}, or using mixed-length vector pruning \cite{tf-mvp}. Significant dataflow and architecture innovations include wafer-scale engines \cite{cerebras}, streaming processors \cite{lpu}, and chiplet-based hybrid systems \cite{cambriconllm}.

\subsection{Processing-in-Memory (PIM) Acceleration}


Concurrently, PIM has been explored through hybrid systems combining NPUs \cite{neupim}, hybrid ReRAM/3D-SRAM \cite{guo-rram}, or 3D mixed-signal integration \cite{h3d}. Advanced packaging is also leveraged via 3D-stacked DRAM \cite{Sharda25}, hybrid-bonding \cite{h2llm}, stacking on DDR5 \cite{pim-ai}, or CXL-based memory expansion \cite{cxl-pnm}. ROM based accelerators have been proposed to achieve extreme memory density by storing the quantized base model on-chip \cite{wang2025roma}.

\section{Conclusion}

This paper confronts the critical "memory wall" challenge that impedes the deployment of Large Language Models on power-constrained edge devices.
While existing solutions like PIM or 3D-stacked memory have been explored, they often fail to simultaneously deliver the extreme density, massive bandwidth, and low power required for real-time edge intelligence.

To address this, 
we introduce TOM, a Ternary-Oriented accelerator featuring a hybrid ROM-SRAM architecture.
Our work makes three primary contributions.
First, we propose a sparsity-aware ROM architecture that synthesizes ternary-quantized weights as standard-cell logic, eliminating the area cost of zero-bits to achieve state-of-the-art storage density.
Second, we develop a distributed processing architecture that co-locates these dense ROM banks with computation, unlocking enormous parallel bandwidth and minimizing data movement.
Third, we implement a workload-aware dynamic power gating scheme that leverages the logic-based nature of our ROM to power down inactive weight banks, drastically reducing static power consumption without any performance penalty.

Our evaluation demonstrates that \aname achieves an end-to-end speedup of up to 63.7x and a throughput of approximately 3306 Tokens Per Second when compared to a high-end NVIDIA A100 GPU.
By fundamentally rethinking on-chip memory for LLMs, \aname demonstrates a viable and highly efficient path toward enabling the next generation of real-time, high-performance AI on edge devices.


\bibliographystyle{IEEEtranS}
\bibliography{refs}

\end{document}